# Effect of Al$_2$O$_3$ on the operation of SiN$_X$-based MIS RRAMs

A.E. Mavropoulis[a], N. Vasileiadis[a,b], P. Normand[a], C. Theodorou[c], G. Ch. Sirakoulis[b], S. Kim[d], P. Dimitrakis[a]

[a] Institute of Nanoscience and Nanotechnology, NCSR "Demokritos", Ag. Paraskevi 15341, Greece
[b] Department of Electrical and Computer Engineering, Democritus University of Thrace, Xanthi 67100, Greece
[c] Univ. Grenoble Alpes, Univ. Savoie Mont Blanc, CNRS, Grenoble INP, IMEP-LAHC, 38000 Grenoble, France
[d] Division of Electronics and Electrical Engineering, Dongguk University, Seoul 04620, South Korea

**Abstract**

The role of a 3 nm Al$_2$O$_3$ layer on top of stoichiometric LPCVD SiN$_x$ MIS RRAM cells is investigated by using various electrical characterization techniques. The conductive filament formation is explained, and a compact model is used to fit the current-voltage curves and find its evolution during each operation cycle. The conduction in SiN$_x$ is also studied.

**1. Introduction**

A large variety of resistive random-access memory (RRAM) technologies are prominent. Nevertheless, a few fulfill the requirements for CMOS integration and meet the commercialization standards. SiN$_x$, which is widely used for charge-trapping nonvolatile memories in Flash SONOS [1, 2, 3] and Vertical CT-NVMs [4], is found to exhibit competitive resistance switching (RS) properties and attractive SiN$_x$-based RRAM devices have been recently demonstrated [5, 6, 7] utilizing a metal-insulator-semiconductor structure. These properties render SiN$_x$ as a promising candidate for neuromorphic, in-memory and edge computing [6, 8, 9], as well as for security applications by creating true random number generators [10, 11]. Moreover, Al$_2$O$_3$ has been used in the past as a buffer layer in RRAMs to improve the RS and the cycle-to-cycle variations [12]. In this context, we investigated the effect of inserting a thin Alumina layer between the top electrode and SiN$_x$.

Silicon wafers were implanted with Phosphorous using 60 keV and a dose of 1×10$^{15}$ cm$^{-2}$ and were then annealed at 950 °C for 20 min in an N$_2$ environment, creating a heavily doped n$^{++}$ (Nd = 1×10$^{20}$ /cm$^3$) substrate. A 7 nm LPCVD SiN$_x$ (x = 1.27) was deposited on the Silicon wafer. After that a 3 nm Al$_2$O$_3$ layer was deposited by MEMS ALD. Finally, the top electrode (TE) was realized using 30 nm Cu and 30 nm Pt to prevent oxidation. The metal contact on the Si bottom electrode (BE) was formed by Al evaporation. A sample without the Al$_2$O$_3$, called hereafter reference, was prepared for comparison using the exact same processing steps mentioned previously. The structure of the fabricated devices is presented in the inset of Figure 1a. The DC electrical characterization of the RRAMs is performed using HP4155 and Tektronix 4200A, and the impedance spectroscopy measurements using HP4284 and Zurich Instruments MFIA.

**2. Electrical characterization**

To start with, current-voltage sweeps were performed using different compliance currents (I$_{CC}$) during the SET process. The voltage was applied on the TE and the BE was always grounded. Characteristic I-V curves are presented in Figure 1b. The curves of the reference and the Al$_2$O$_3$/SiN$_x$ devices are similar in terms of current levels for the HRS and LRS. However, the major difference resides in the SET/RESET voltages, which are significantly higher for the Al$_2$O$_3$/SiN$_x$ samples. This is proven by statistically analysing the SET/RESET voltages for both samples (Figure 1c). The standard deviation to mean ratio (σ/μ) is calculated, which is indicative of the variation of these voltages. The addition of Al$_2$O$_3$ slightly reduced the high resistance state (HRS) variability. Nevertheless, the low resistance state (LRS) variability increased. It can also be seen that there is a 0.9 V difference in the mean SET voltage between the samples (Table 1), which can be attributed to the voltage drop on the Al$_2$O$_3$. By calculating the band diagram (Figure 1c) using the Boise State University Band Diagram Program [13], it is found that the voltage drop on the 3 nm Al$_2$O$_3$ layer when +5 V is applied to the top electrode (TE) is 0.9 V, which corresponds exactly to the value derived from the mean SET voltages calculated before. The increased SET voltage indicates that that the addition of the Al$_2$O$_3$ only added a voltage drop and the resistance switching phenomena take place exclusively inside the SiN$_x$.

Table 1 Mean SET/RESET voltages and standard deviation.

| Sample | V$_{SET}$ | | | V$_{RESET}$ | | |
|---|---|---|---|---|---|---|
| | μ | σ | σ/μ | μ | σ | σ/μ |
| Al$_2$O$_3$/SiN$_x$ | 4.54 | 0.67 | 0.15 | -4.22 | 0.43 | 0.10 |
| Reference | 3.61 | 0.19 | 0.05 | -2.92 | 0.43 | 0.15 |

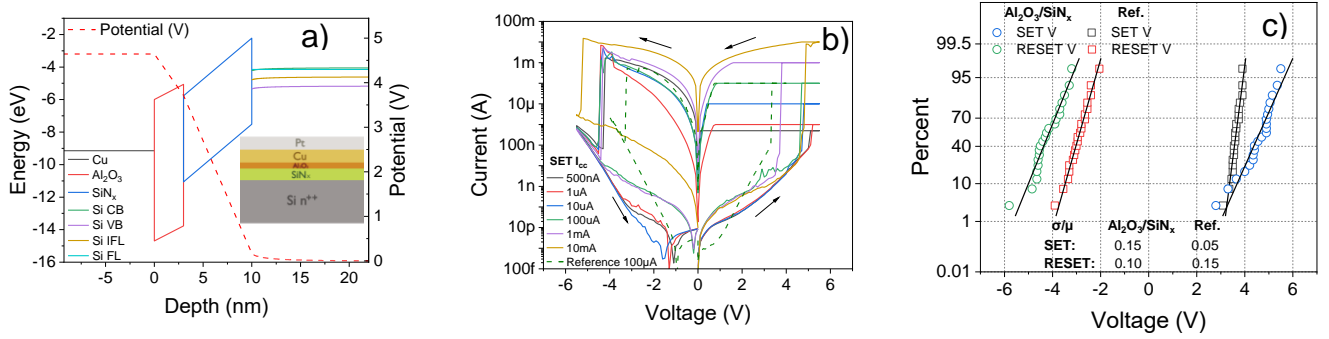

Figure 1 a) Ideal band diagram of $Al_2O_3/SiN_x$ under +5 V bias on the TE (left y axis) and potential across the dielectric (right y axis), b) I-V sweeps with different $I_{CC}$ and c) SET/RESET voltage statistics for the $Al_2O_3/SiN_x$ and reference samples.

After that, multiple SET/RESET cycles were performed using voltage sweeps with the same conditions ($I_{CC}$ = 100 μA, 5.5 V for SET and -7 V for RESET) and the current after each SET/RESET operation was measured at 0.5 V. Typically, the samples with $Al_2O_3$ were able to perform more than 100 cycles, with further improvement for optimization by fine tuning the SET/RESET voltages. Endurance could also be improved by changing the substrate to SOI, where the self-compliance property of the $SiN_x$ memristors has been demonstrated [14]. Furthermore, it should be emphasized that voltage sweep measurements degrade device performance severely compared to pulsed or ramped voltage operation [15]. The memory window evolution, which is the ratio between LRS/HRS, is shown in Figure 2a. Initially, this window is $10^4$ and after ~20 cycles it stabilizes to $10^5$. This is a similar behavior to the one studied before on the reference sample [16], where the HRS read current decreased in the first 30 cycles. This can be explained by the reduction of the conductive filament gap of the LRS after each cycle. This conductive filament (CF) is formed from traps, specifically Nitrogen vacancies which result in Silicon dangling bonds and allow current to flow through them under electric field [17, 18]. The fact that the CF is created by Nitrogen vacancies means that it is possible to simulate the I-V curves using the compact model of Chen P.-Y [19], which is based on oxygen vacancies in an oxide that act as electron hopping sites. The equation that describes the relationship between current and voltage is:

$$I = I_0 \exp\left(-\frac{g}{g_0}\right) \sinh\left(\frac{V}{V_0}\right) \qquad (1)$$

where g is the gap distance between the tip of the CF and the TE and $I_0$, $g_0$, $V_0$ are fitting parameters, denoting the current, the gap and the voltage. According to [19], the parameters $I_0$, $g_0$ and $V_0$ do not have a specific physical meaning and are used to describe the non-linearity of the I-V characteristics. It is found that the I-V curves of our samples fit very well with this model and the evolution of the CF every cycle can be calculated. The CF exponentially decays until cycle 30 as seen in Figure 2b. This means that the model can be successfully used to simulate our devices in circuit level designs. $I_0$ and $g_0$ are the same during the fitting of the I-V for every cycle ($I_0$ = 25.4 μA, $g_0$ = 0.275 nm). However, $V_0$ changes from cycle to cycle, as shown in Figure 2c, without exhibiting any discernable dependance. In addition, the increased LRS variability after 30 cycles (Figure 2a) is mainly attributed to the stochastic number of conductive paths, the stochastic process of ion movement as well as the variation of the gap, g. These processes have stronger impact due to the degradation of the cell's life. The ambient temperature increase will potentially result in higher currents (depending on the conduction mechanism, e.g., Ohmic, Poole-Frenkel, TAT), which will lead to greater variation of the HRS/LRS and the cell endurance will be reduced [20]. The local temperature increase results in higher energy ion movement and thicker filaments during switching. Thus, LRS exhibits greater variation due to the stochastic variation of the gap and is reduced due to the larger diameter of the conductive filaments. Furthermore, LRS can get stuck at a specific value at high temperatures because the kinetic energy of the ions moving towards the TE causes chemical reactions and so, the TE is chemically modified [21, 22].

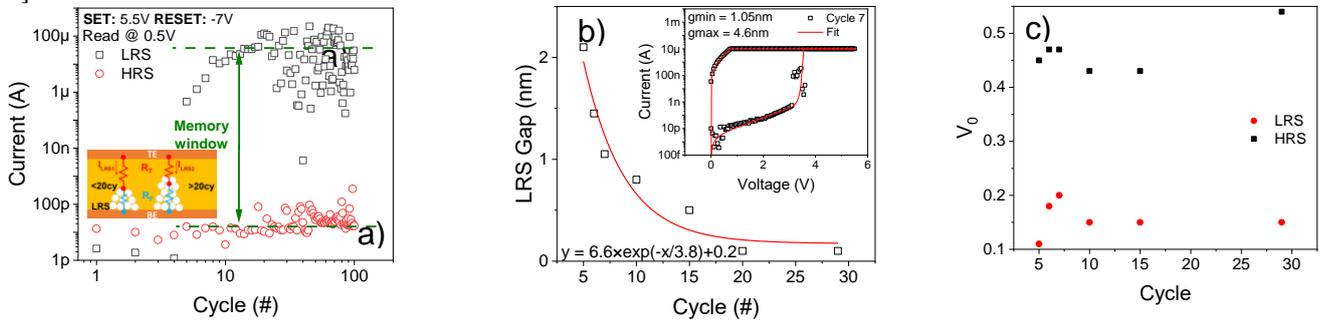

Figure 2 Multiple operation cycles with the same SET/RESET conditions a) HRS/LRS at 0.5 V, b) CF evolution at the first 30 cycles (inset depicts a typical fit in the $7^{th}$ cycle) and c) $V_0$ fitting parameter evolution for each cycle.

Moreover, impedance spectroscopy measurements were performed for pristine and cycled devices, i.e., after the SET/RESET sweeps using different SET $I_{CC}$, utilizing an AC small signal of 25 mV and a DC bias of 0.1 V on the TE. The Nyquist plots for $Al_2O_3/SiN_x$ devices at LRS form a semicircle (Figure 3a) indicative of an equivalent circuit consisting of a resistor ($R_p$) with a capacitor ($C_p$) in parallel and a resistor in series ($R_s$). Physically, $R_p$ and $C_p$ correspond to the resistance of the conductive paths formed during SET and the capacitance of the remaining insulating (no-switched) material region, respectively. $R_p$ decreases with increasing SET $I_{CC}$ (Table 2), resulting in a more conductive CF. The dielectric constant ($\varepsilon' = Re(\varepsilon^*) = \varepsilon$) was also extracted (Figure

3b) for pristine samples, and it was found to be significantly higher (~7.7) than the reference sample (~5.5), which is expected due to the addition of the 3 nm $Al_2O_3$, with a reported dielectric constant of 6.7 [23]. Furthermore, the AC conductance (σ') was calculated (Figure 3c) by subtracting the DC part from the measurements and it becomes clear that σ' varies as ~$f^s$. The values of exponent s range from 1.59 to 1.67 for the LRS and 1.69 for the HRS and denote that the conduction in $SiN_x$ is mainly governed by trap-to-trap tunneling mechanisms (s is close to 2) [24]. The equations

$$\varepsilon(\omega) = \frac{1}{i\omega C_0 Z(\omega)} \quad (2)$$

and

$$\sigma(\omega) = i\omega\varepsilon_0\varepsilon(\omega) \quad (3)$$

were used to calculate the dielectric constant and the conductance for the impedance measurements, where ω the angular frequency, $\varepsilon_0$ the permittivity of free space and $C_0$ the geometrical capacitance ($C_0 = A\varepsilon_0/d$, d: thickness of the dielectric, A: area of the dielectric) [25].

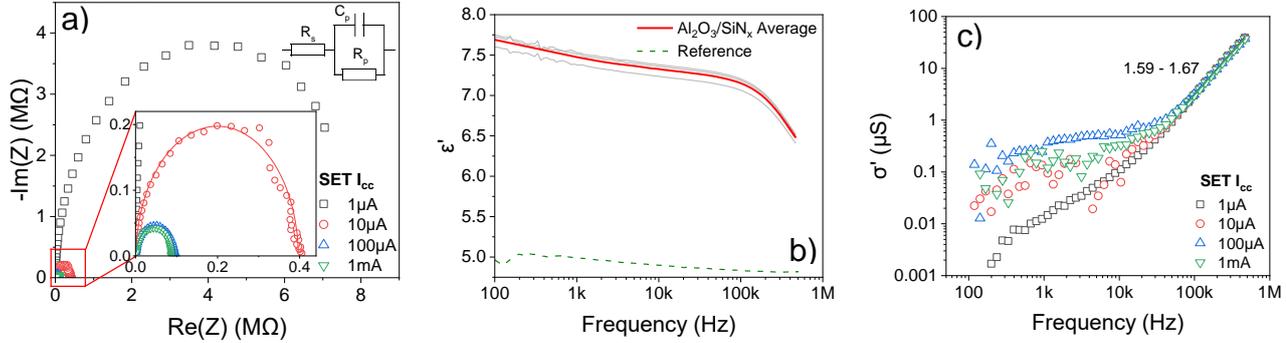

Figure 3 a) Nyquist plots of $Al_2O_3/SiN_X$ devices SET with different $I_{CC}$, b) dielectric constant of pristine sample and c) AC conductance in the LRS after SET with different $I_{CC}$.

Table 2 Modelling parameters for Nyquist plots at LRS.

| SET $I_{CC}$ | $C_p$ (pF) | $R_p$ (kΩ) |
|---|---|---|
| 10 μA | 62.1 | 395 |
| 100 μA | 61.3 | 94.8 |
| 1 mA | 61.1 | 123.5 |

## 3. Conclusions

The switching characteristics of memristors with and without $Al_2O_3$ reveal that the formed filament at LRS is probably due to the formation of nitrogen vacancies and not due to Cu diffusion in the $SiN_x$. It is concluded that these devices can be SET at different resistance levels by varying the $I_{CC}$. Moreover, the SET/RESET voltages are higher compared to the reference sample, corresponding to the voltage drop on the $Al_2O_3$. This device can perform multiple operation cycles, with the memory window increasing in the first twenty, which according to a compact model fit of the I-V curves, is attributed to the decrease of the LRS gap between the CF and the TE. The filamentary conduction at LRS is also proved by the applied model. In addition, impedance spectroscopy measurements revealed that the conduction in $SiN_x$ is mostly governed by trap-to-trap tunneling mechanisms.

## Acknowledgements

This work was financially supported by the research project "LIMA-chip" (Proj.No. 2748) of the Hellenic Foundation for Research and Innovation (HFRI).